\documentclass[10pt, conference, twocolumn, letterpaper]{IEEEtran}
\usepackage[utf8x]{inputenc}
\usepackage[T1]{fontenc}
\usepackage{textcomp}
\usepackage{amsmath}
\usepackage{amsthm}
\usepackage{amsfonts}
\usepackage{mathtools}
\usepackage{amssymb}
\usepackage{graphicx}
\usepackage{booktabs}
\usepackage{tabularx}
\usepackage{makecell}
\usepackage{chngcntr}
\usepackage{subfigure}

\usepackage{algpseudocode}
\usepackage{algorithm} 
\usepackage{algorithmicx}
\usepackage{multirow}
\usepackage{paralist}
\usepackage{url}

\counterwithin{paragraph}{section}

\usepackage[capitalise]{cleveref} 
\newtheorem{observation}{Observation}
\newtheorem{theorem}{Theorem}

\newcolumntype{L}{>{\raggedright\arraybackslash}X}
\newcolumntype{C}{>{\centering\arraybackslash}X}

\algblockdefx[DOWHILE]{Do}{}{\textbf{do}}{}
\algblockdefx[DOWHILE]{}{While}[1]{\textbf{while} #1}

\title{Securing Virtual Network Function Placement with High Availability Guarantees}

\author{
\IEEEauthorblockN{Marco Casazza\IEEEauthorrefmark{1}, Pierre Fouilhoux\IEEEauthorrefmark{2}, Mathieu Bouet\IEEEauthorrefmark{3}, and Stefano Secci\IEEEauthorrefmark{2}}
\IEEEauthorblockA{\IEEEauthorrefmark{1}Università degli Studi di Milano, Dipartimento di Informatica, Italy. Email: marco.casazza@unimi.it}
\IEEEauthorblockA{\IEEEauthorrefmark{2}Sorbonne Universités, UPMC Univ Paris 06, UMR 7606, LIP6, France. Email: \{pierre.fouilhoux,stefano.secci\}@upmc.fr}
\IEEEauthorblockA{\IEEEauthorrefmark{3}Thales Communications \& Security, France. Email: mathieu.bouet@thalesgroup.com}
}

\begin{document}

\maketitle

\begin{abstract}
Virtual Network Functions as a Service (VNFaaS) is currently under attentive study by telecommunications and cloud stakeholders as a promising business and technical direction consisting of providing network functions as a service on a cloud (NFV Infrastructure), instead of delivering standalone network appliances, in order to provide higher scalability and reduce maintenance costs.
However, the functioning of such NFVI hosting the VNFs is fundamental for all the services and applications running on top of it, forcing to guarantee a high availability level.
Indeed the availability of an VNFaaS relies on the failure rate of its single components, namely the servers, the virtualization software, and the communication network.
The proper assignment of the virtual machines implementing network functions to NFVI servers and their protection is essential to guarantee high availability.
We model the High Availability Virtual Network Function Placement (HA-VNFP) as the problem of finding the best assignment of virtual machines to servers guaranteeing protection by replication. We propose a probabilistic approach to measure the real availability of a system and design both efficient and effective algorithms that can be used by stakeholders for both online and offline planning.
\end{abstract}

\section{Background}

A recent trend in computer networks and cloud computing is to virtualize network functions, in order to provide higher scalability, reducing maintenance costs, and increasing reliability of network services.
Virtual Network Functions as a Service (VNFaaS) is currently under attentive study by telecommunications and cloud stakeholders, as a promising business and technical direction consisting of providing network functions (i.e., firewall, intrusion detection, caching, gateways...) as a Service instead of delivering standalone network appliances. While legacy network services are usually implemented by means of highly reliable hardware specifically built for a single purpose middlebox, VNFaaS moves such services to a virtualized environment~\cite{clickos}, named \emph{NFV Infrastructure (NFVI)} and based on commercial-off-the-shelf hardware~\cite{Mo2015}.

Services implementing network functions are called \emph{Virtual Network Functions (VNFs)}.
One of the open issues for NFVI design is indeed to guarantee high levels of VNF availability \cite{Mijumbi2016}, i.e., the probability that the network function is working at a given time. In other words, a higher availability corresponds to a smaller downtime of the system, and it is required to satisfy stringent \emph{Service Level Agreements (SLA)}. 
Failures may result in a temporary unavailability of the services, but while in other contexts it may be tolerable, in NFVI network outages are not acceptable, since the failure of a single VNF can induce the failure of all the overlying services  \cite{nfv-resiliency}.
To achieve high availability, backup VNFs can be placed into the NFVI, acting as replicas of the running VNFs, so that when the latter fail, the load is rerouted to the former.
However, not all VNFs are equal ones: 
the software implementing a network function of the server where a VNF is running may be more prone to errors than others, influencing the availability of the overall infrastructure.
Also, client requests may be routed via different network paths, with different availability performance.
Therefore to guarantee high levels of availability it is important not only to increase the number of replicas placed on an NFVI, but  it is also crucial to select where they are placed and which requests they serve.

In this context, we study and model the \emph{High Availability Virtual Network Function Placement (HA-VNFP)} problem, that is the problem of placing VNFs on an NFVI in order to serve a given set of clients requests guaranteeing high availability.
Our contribution consist of:
\begin{inparaenum}[(a)]
\item a quantitative probabilistic model
to measure the expected availability of VNF placement;
\item a proof that the problem is $\mathcal{NP}$-hard and that it belongs to the class of nonlinear optimization problems;
\item a linear mathematical programming formulation that can be solved to optimality for instances of limited size;
\item a \emph{Variable Neighborhood Search (VNS)} heuristic for both online and offline planning;
\item an extensive simulation campaign, and algorithm integration in a Decision Support System (DSS) tool~\cite{HANFVsw}.
\end{inparaenum}

The paper is organized as follows: 
in \cref{sec:description} we present the HA-VNFP, and in \cref{sec:literature} we briefly describe previous works on VM/VNF placement in cloud/NFVI  systems. 
In \cref{sec:modelling} we formally describe the optimization problem and propose a linearization of the problem and a mathematical programming formulation that can solve it to optimality.
In \cref{sec:algorithms} we describe our heuristic methodologies, which are then tested in an extensive simulation campaign in \cref{sec:simulation}. We briefly conclude in \cref{sec:conclusion}.

\section{High Availability VNF Placement}\label{sec:description}

\begin{figure}[ht]
\centering
\includegraphics[width=0.9\columnwidth]{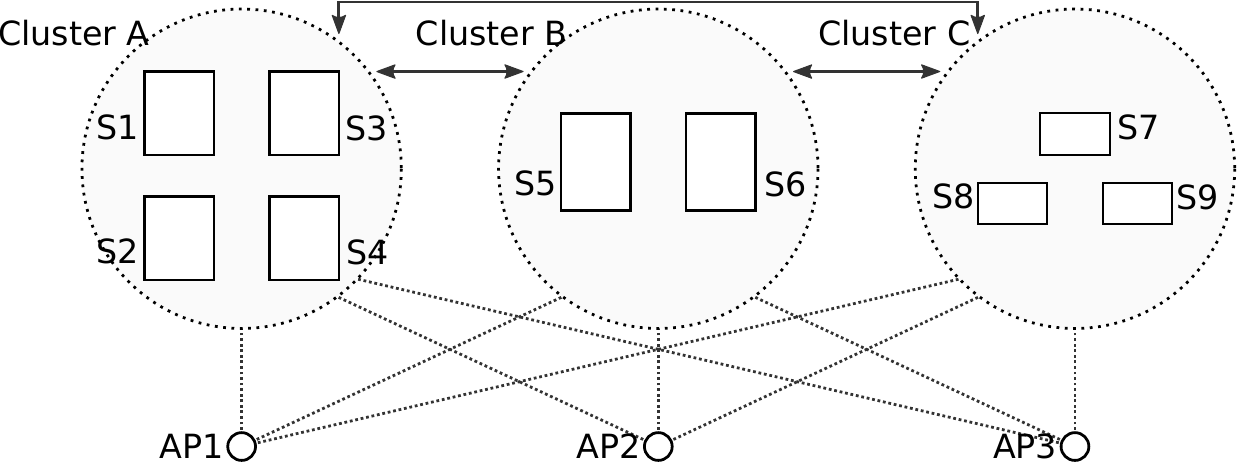}
\caption{Abstract representation of an NFVI: each server is depicted as a white box whom height represents the amount of available resources. Clusters are connected together to allow synchronization operations. Access network is represented by access points connecting clusters to the external network.}
\label{figure:nfv-abstract-infrastructure}
\end{figure}

We consider an NFVI with several geo-distributed datacenters or \emph{clusters} (see \cref{figure:nfv-abstract-infrastructure}). Each cluster consists of a set of heterogeneous servers with limited available computing resources. 
Several instances of the same VNF type can be placed on the NFVI but on different servers.
Each VNF instance can be assigned to a single server allocating some of its computing resources.
Indeed each server has a limited amount of computing resources that cannot be exceeded.

A network connects together all servers of the NFVI: we suppose that the communication links inside a cluster are significantly more reliable than those between servers in different clusters. 
An access network with multiple access points connects servers to clients. Links connecting access points to clusters can differ in the availability level, depending on the type of the connection or the distance from the cluster.

In this article, we assume that the total amount of resources and network capacities are sufficient to manage the expected client requests at any time. However assignment decisions may artificially produce congestion over the servers. We analyze how to find assignments providing a trade-off between NFVI availability and system congestion.

We are given an estimation of the expected client VNF requests, each characterized by a computing resource demand. 
An assigned request consumes part of the resources reserved by a VNF instance.
Indeed the consumed resources must not exceed the reserved ones.

Requests can be assigned using two different policies:
\emph{demand load balancing} and \emph{split demand load balancing}. In the former, a client request is always fully assigned to a single server, while in the latter it may be split among different ones. Splitting a request also splits proportionally its demand of computing resources. Indeed, when a demand is split it relies on the availabilities of many VNF instances, decreasing the expected availability of the service, but increasing the chance of finding a feasible assignment in case of congestion.

\begin{figure}[ht]
\centering
\includegraphics[width=\columnwidth]{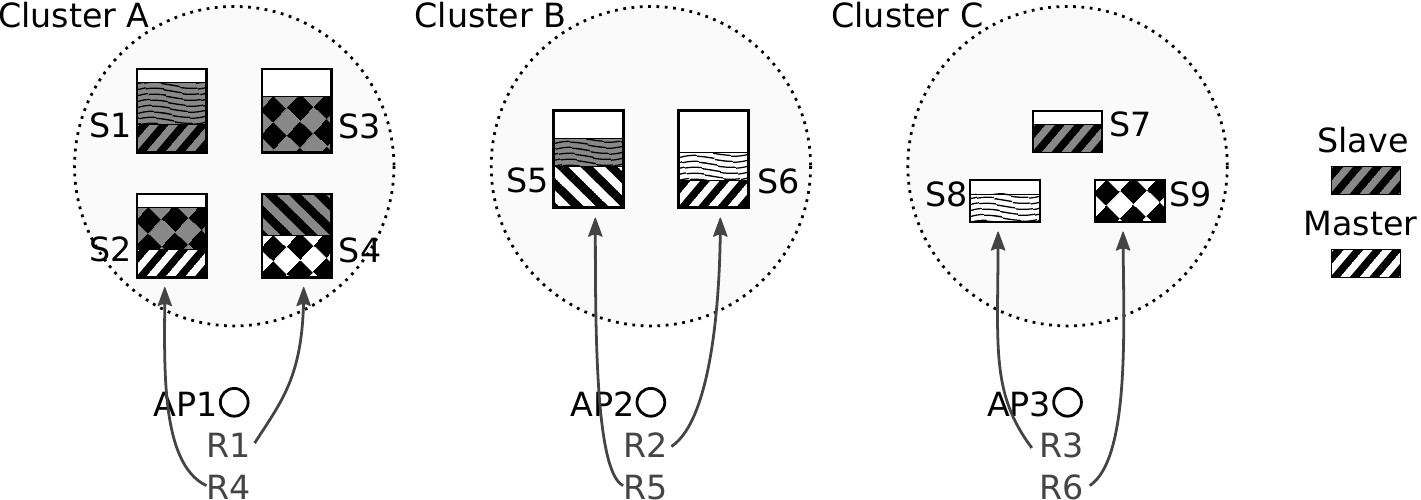}
\caption{Abstract representation of VNFs placement on a 3-cluster NFVI. Each box is a VNF instance. Instances running the same VNF type have the same pattern. VNFs with a gray background are slaves placed as protection for the masters. Sets of requests are routed from access points and assigned to VNFs running the requested function.}
\label{figure:nfv-abstract-placement}
\end{figure}

We suppose a multi-failure environment in which VNFs, servers, clusters, and networks may fail together.
Our aim is to improve the VNF availability by replicating instances on the NFVI. We distinguish between \emph{master} and \emph{slave} VNFs: the former are active VNF instances, while the latter are idle until masters fail.
An example of VNF placement is depicted in \cref{figure:nfv-abstract-placement}.

Each master may be \emph{protected} by many slaves - we assume in this article that a slave can protect only a master, must be placed on a different server, and must allocate at least the same amount of computing resources of its master.

\begin{figure}[ht]
\centering
\includegraphics[width=0.55\columnwidth]{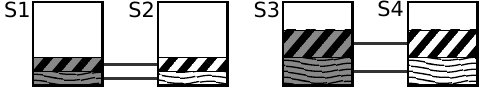}
\caption{Example of protection: master VNFs are running on servers $S2$ and $S4$, while slave are running on $S1$ and $S3$. Each link between VNF instances represents the connection between a master and its slave.}
\label{figure:nfv-dp}
\end{figure}

Each master periodically saves and sends its state to its slaves, e.g. using technologies such as the one presented in~\cite{Cully08}, in such a way that the latter has always an updated state and can consistently restore the computation
in case of failure of the former.
We suppose that if a master is unreachable, a searching process is started in order to find a slave that can complete the computation. If the searching process fails and all slaves are unreachable, then the service is considered unavailable.
A representation of VNF protection is in \cref{figure:nfv-dp}.

\section{Related works}\label{sec:literature}
Even if VM and VNF resource placement in cloud systems is a recent area of research (see \cite{Jennings2014} for a high-level comprehensive study), however
there already exists orchestrators that are driven by optimization algorithms for the placement, such as~\cite{Riera2016}.
We now present few works in literature studying 
the optimization problems that arises in this context.

\paragraph{Placement of Virtual Machines}
\cite{Meng10} studies the problem of placing VMs in datacenters minimizing the average latency of VM-to-VM communications. Such a problem is $\mathcal{NP}$-hard and falls into the category of \emph{Quadratic Assignment Problems}. The authors provide a polynomial time heuristic algorithm solving the problem in a \emph{"divide et impera"} fashion.
In \cite{Alicherry12} the authors deal with the problem of placing VMs in geo-distributed clouds minimizing the inter-VM communication delays. They decompose the problem in subproblems that they solve heuristically. They also prove that, under certain conditions, one of the subproblems can be solved to optimality in polynomial time.
\cite{Biran12} studies the VM placement problem  minimizing the maximum ratio of the demand and the capacity across all cuts in the network, in order to absorb unpredictable traffic burst.
The authors provide two different heuristics to solve the problem in reasonable computing time.

\paragraph{Placement of Virtual Network Functions}
\cite{Basta14} applies NFV to LTE mobile core gateways proposing the problem of placing VNFs in datacenters satisfying all client requests and latency constraints while minimizing the overall network load.
Instead, in~\cite{Cohen15}  the objective function requires to minimize the total system cost, comprising the setup and link costs. 
\cite{Luizelli15} introduces the VNF orchestration  problem of placing VNFs and routing client requests through a chain of VNFs.
The authors minimize the setup costs while satisfying all client demands. They propose both an ILP and a heuristic to solve such problem.
Also \cite{Addis15} considers the VNF orchestration problem with VNF switching piece-wise linear latency function and bit-rate compression and decompression operations. Two different objective functions are studied: one minimizing costs and one balancing the network usage. 

\paragraph{Placement with protection}
In \cite{Bin11} VMs are placed with a protection guaranteeing $k$-resiliency, that is at least $k$ slaves for each VM. The authors propose an integer formulation that they solve by means of constraint programming.
In~\cite{Israel13} the recovery problem of a cloud system is considered where slaves are usually turned off to reduce energy consumption but can be turned on in advance to reduce the recovery time.
The authors propose a bicriteria approximation algorithm and a greedy heuristic.
In~\cite{Zhu14} the authors solve a problem where links connecting datacenters may fail, and a star connection between VMs must be found minimizing the probability of failure. The authors propose an exact and a greedy algorithms to solve both small and large instances, respectively.
Within disaster-resilient VM placement, \cite{Couto15} proposes a protection scheme in which for each master a slave is selected on a different datacenter, enforcing also path protection.
In~\cite{Alameddine2016} the authors solve the problem of placing slaves for a given set of master VMs without exceeding neither servers nor link capacities. Their heuristic approaches decompose the problems in two parts: the first allocating slaves, and the second defining protection relationships.

In a recent work~\cite{Yang2016}, the authors model the VM availability by means of a probabilistic approach and solve the placement problem over a set of servers by means of a nonlinear mathematical formulation and greedy heuristics. This is the only work offering an estimation of the availability of the system. However, it considers only the availability of the servers, while in our problem we address a more generic scenario: when datacenters are geo-distributed, a client request shall be assigned to the closest datacenter, since longer connections may have a higher failure rate. 
Therefore, the source of the client requests may affect the placement of the VNFs on the NFVI, and must be taken into account in the optimization process and in the estimation of the availability. 

\section{Modeling}
\label{sec:modelling}
In the following we propose a formal definition to the HA-VNFP and a mathematical programming formulation.

\paragraph{Clusters and servers}
We are given the set of clusters $C$ and the set of servers $S$.
Each server $s$ belongs to a cluster $c_s$, and we define as $S_c \subseteq S$ the set of servers of cluster $c$.
We represent the usual distinct types of computing resources (CPU, RAM, ... ) of server $s \in S$ by the same global amount $q_s \in \mathbb{R}_+$ of available resources. 

\paragraph{Virtual Network Functions} 
A set $F$ of VNF types is given. 
Each VNF instance runs on a single server. Each server can host multiple VNF instances, but at most one master for each type.

\paragraph{Networks} 
An inter-cluster network allows synchronization between clusters, while an access network connects clusters to a set of access points $P$.
We are given sets $L_{C}$ and $L_P$ of logical links $(c',c'') \in L_C$ connecting clusters $c',c'' \in C$, and logical links $(c,p) \in L_P$ connecting cluster $c \in C$ to access point $p \in P$, respectively.

\paragraph{Clients requests} 
A set of clients requests $R$ is given. Each request $r \in R$ is a tuple $(f_r, P_r, d_r)$ of the requested VNF type $f_r \in F$, a subset of available access points $P_r \subseteq P$, and the resources demand $d_r \in \mathbb{R}_{+}$.

\paragraph{Availability} 
Taking into account explicit availability in NFVI design becomes necessary to ensure SLAs~\cite{ nfv-resiliency,Mo2015}.
We suppose that the availabilities of each component (server, cluster, VNF, link) are given (see \cref{tab:data}), each corresponding to the probability that a component is working.

\paragraph{Objective function}
All clients requests must be assigned to servers maximizing the availability of the system, we measure  as the minimum availability among all requests.

\begin{table}[ht]
\caption{Mathematical notation.}
\label{tab:data}
\begin{tabularx}{\columnwidth}{c|L|c|L}
\toprule
$C$ & Set of clusters &
$S$ & Set of servers\\
$S_c$ & Set of servers in cluster $c$ &
$F$ & Set of VNF types\\
$R$ & Set of requests&
$P$ & Set of access points\\
$L_C$ & Set of synchro. links &
$L_P$ & Set of access links\\
$P_r$ & Set of request access points& \\
\midrule
$q_s$ & Capacity of server $s$ & 
$d_r$ & Demand of request $r$\\
$c_s$ & Cluster of server $s$ & 
$f_r$ & VNF of request $r$\\
\midrule
$a_{f}^F$ & Availability of VNF $f$ &
$a_{s}^S$ & Availability of server $s$ \\
$a_{cc'}^{L_C}$ & Availability of synchro. link $(c,c')$ & $a_{cp}^{L_P}$ & Availability of access link $(c,p)$ \\
$a_{c}^C$ & Availability of cluster $c$ &
\\
\bottomrule
\end{tabularx}
\end{table}

\subsection{Computational complexity}
Concerning the assignment of requests, we can prove that:
\begin{observation}\label{theo:feasibility-split}
When demand split is allowed and $\sum_{r \in R} d_r \le \sum_{s \in S} q_s$, HA-VNFP has always a feasible solution that can be found in polynomial time.
\end{observation}
In fact since the requests can be split among servers, the feasibility of an instance can be found applying a Next-Fit greedy algorithm for the Bin Packing Problem with Item Fragmentation (BPPIF) \cite{Meankerman2001}: servers can be seen as bins, while requests as items that must be packed into bins. 
The algorithm iteratively pack items to an open bin. When there is not enough residual capacity, the item is split, the bin is filled and closed, and a new bin is open packing the rest of the item.
When requests can be split, such algorithm produces a feasible solution for the HA-VNFP:
if a request is assigned to a server, then a master VNF serving such a request is allocated on that server too. The Next-Fit algorithm runs in $O(|R|)$ and therefore a feasible solution can be found in polynomial time.
\begin{observation}\label{theo:feasibility-nosplit}
The feasibility of a HA-VNFP instance without demand split is a $\mathcal{NP}$-hard problem.
\end{observation}
Indeed we can see again the feasibility problem as a Bin Packing Problem (BPP). However, without split each item must be fully packed into a single bin.
Therefore, finding a feasible solution is equivalent to the feasibility of a BPP, which is $\mathcal{NP}$-hard, and 
it directly follows that:
\begin{theorem}\label{theo:complexity-nosplit}
The HA-VNFP without demand split is $\mathcal{NP}$-hard.
\end{theorem}
That is, for unsplittable demands, it is $\mathcal{NP}$-hard finding both a feasible solution and the optimum solution. It is less straightforward to also prove that:
\begin{theorem}\label{theo:complexity-split}
The HA-VNFP with demand split is $\mathcal{NP}$-hard.
\end{theorem}
\begin{IEEEproof}
In fact, let us suppose a simple instance where all components (servers, clusters, links, ...) are equal ones and where $\sum_{r \in R} d_r = \sum_{s \in S} q_s$, which means that there will be no slaves in our placement.
The problem can be seen again as a BPPIF in which the objective is to minimize the number of splits of the item that is split the most: in fact, every time a request is split, the availability of the system decreases. In such scenarios the best solution is the one in which no request is split at all - however, if we could solve such a problem in polynomial time, then we could solve also the feasibility problem of a BPP in polynomial time, which instead is $\mathcal{NP}$-hard. Therefore, since we can reduce a feasibility problem of BPP to an instance of BPPIF, and the latter to an instance of HA-VNFP, the HA-VNFP with split is $\mathcal{NP}$-hard.
\end{IEEEproof}

\subsection{Mathematical formulation}
\begin{figure}[ht]
\centering
\includegraphics[width=0.9\columnwidth]{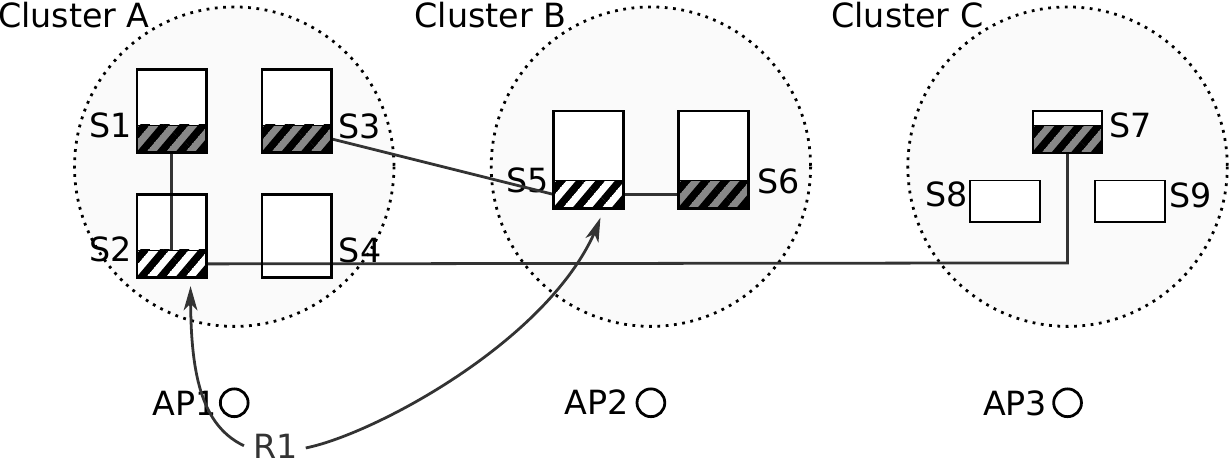}
\caption{Example of assignment configuration $\gamma = \{(S2, \{S1, S2, S7\}), (S5,\{S3, S5, S6\}) \}$, where request $R1$ is split and assigned to two different master VNFs on servers $S2$ and $S5$. Both masters have slaves: the master VNF on server $S2$ has slaves on servers $S1$ and $S7$, while the one on server $S5$ has slaves on servers $S3$ and $S6$.}
\label{fig:assignment-configurations}
\end{figure}

In the following we propose a mathematical programming formulation of HA-VNFP starting from the definition of the set of the solutions:
a \emph{request assignment} $\omega$ is a pair $(s, S_p)$ indicating the subset of servers $S_p \subseteq S $ running either the master or the slaves of a VNF instance, and the server $s \in S_p$ where the master is placed.
We also define $\Omega = \{(s, S_p) \mid S_p \subseteq S, s \in S_p\}$ as the set of all request assignments.
An \emph{assignment configuration} $\gamma$ (see \cref{fig:assignment-configurations}) is a set of all request assignments $\omega$ for all the fragments of a request.
We define as $\Gamma$ the set of all assignment configurations $\gamma$, that is $
\Gamma = \{ \gamma \in 2^\Omega \mid s' \neq s'', \forall (s',\omega'), (s'', \omega'') \in \gamma\}.
$

\paragraph{Availability computation}
We compute the NFVI availability for a request $r$ by means of a probabilistic approach \cite{Birolini2004,iec-reliability}. 
Given a cluster and a set of access points, $a^{L_P}(c,P)$ is the function computing the probability that at least one of the access links is working:
$
a^{L_P}(c,P) = 1 - ( \prod_{p \in P} 1 - a_{cp}^{L_P} ).
$
Given a VNF and a set of servers, $a^{S}(f,S)$ is the function computing the probability that at least one instance of VNF is working:
$
a^{S}(f,S) = 1 - ( \prod_{s \in S} 1 - a_{f}^F \cdot a_s^S ).
$
Given a request $r$ and a request assignment $\omega = (s, S_p)$, $a(r, \omega)$ is the function computing the probability that at least one of the instances of $\omega$ is working:
\begin{multline*}
a(r, \omega)= 1 
- [ \underbrace{( 1 - a^{L_P}(c_s, P_r) \cdot a_{c_s}^C \cdot a^S(f_r, S_p \cap S_{c_s}))}_{\mbox{availability of the cluster containing master}} \cdot  \\
\cdot \prod_{c \in C \setminus \{c_s\}} \underbrace{( 1 -a^{L_P}(c, P_r) \cdot a_{c}^C \cdot a_{c_sc}^{L_C} \cdot a^S(f_r, S_p \cap S_c)}_{\mbox{availability of cluster containing only slaves}} ]
\end{multline*}
When a request $r$ is split,
we compute its availability $a(r, \gamma)$ as the probability that all 
of its parts succeed:
$
a(r, \gamma) = \prod_{(s, \omega) \in \gamma} a(r,\omega).
$
We remark that such formula 
is nonlinear and produces a Integer Nonlinear Programming formulation which cannot be solved by common integer solvers like CPLEX.
Therefore we propose a MIP linearization of such nonlinear formulation in which for each assignment configuration $\gamma \in \Gamma$ we have a binary variable stating if such configuration is selected in the solution.
\paragraph{Variables}
The following variables are needed:
\begin{align*}
x_{rs} : & \mbox{ fraction of request $r$ assigned to server $s$}\\
z_{r\gamma} = &  \left\{ \begin{aligned}
1 , & \mbox{ if $\gamma$ is active for request }r\\
0, & \mbox{ otherwise}
\end{aligned} \right.\\
u_{fs} : & \mbox{ resources consumed by VNF $f$ on $s$}\\
v_{fss'} : & \mbox{ resources consumed by slave on server $s'$ }\\
\mathcal{A}_{min} : & \mbox{ minimum availability}
\end{align*}

\paragraph{Model} HA-VNFP can be modeled as follows:
{
\allowdisplaybreaks
\begin{align}
\max \enspace & \mathcal{A}_{min}\\
\mbox{s.t.}& \sum_{s \in S} x_{rs} = 1 & \substack{\forall r \in R }\label{model:assignment}\\
& x_{rs} \le \sum_{\substack{\gamma \in \Gamma\\\exists (s, \omega) \in \gamma}} z_{r\gamma} & \substack{\forall r \in R,s \in S }\label{model:fraction}\\
&  \sum_{\substack{r \in R \\ f_r = f}} d_{r} \cdot x_{rs}  \le u_{fs} & \substack{\forall f \in F, s \in S} \label{model:vnf-resources}\\
& u_{f_rs} + q_{s} \cdot \sum_{\mathclap{\substack{\gamma \in \Gamma\\\exists (s,\omega) \in \gamma \mid s' \in \omega}}} z_{r\gamma} \le v_{f_rss'}+ q_{s}  & \substack{\forall r \in R, s,s' \in S}\label{model:backup-resources}\\
& \sum_{f \in F} u_{fs} + \sum_{\mathclap{s' \in S}} v_{fs's} \le q_{s} &  \substack{\forall s \in S}\label{model:capacity}\\
& \sum_{\gamma \in \Gamma } z_{r\gamma} \le 1 &\substack{ \forall r \in R}\label{model:max-gamma}\\
& \sum_{\gamma \in \Gamma} a(r,\gamma) \cdot z_{r\gamma}  \ge \mathcal{A}_{min} & \substack{ \forall r \in R }\label{model:min-availability}
\end{align}
}
Constraints \eqref{model:assignment} and \eqref{model:fraction} ensure that each request is fully assigned and selects an assignment configuration, respectively.
Constraints \eqref{model:vnf-resources} and \eqref{model:backup-resources} set the allocated resources of masters and slaves, respectively.
Constraints \eqref{model:capacity} ensure that servers capacities are not exceeded.
Constraints \eqref{model:max-gamma} impose that at most one assignment configuration is selected for each request.
Constraints \eqref{model:min-availability} compute the minimum availability.
Our formulation can model both the HA-VNFP with and without split: in fact by simply setting $|\gamma| = 1$ for each configuration $\gamma$ we forbid configurations splitting a request.

\section{Heuristics}\label{sec:algorithms}
Solving HA-VNFP as a MIP using an integer solver works only for small NFVI,
since the number of variables is exponential w.r.t the size of the instances.
Therefore we propose two different heuristic approaches for HA-VNFP: the first is an adaptation of well-known greedy policies for the BPP that will serve as comparison, while the second is a \emph{Variable Neighborhood Search} heuristic using different algorithmic operators to explore the neighborhood of a starting point.

\subsection{Greedy heuristics}\label{sec:greedy}
Most of the heuristics for the placement of VMs or VNFs are based on a greedy approach, and
BPP heuristics are often exploited to obtain suitable algorithms for the placement, such as in \cite{Silva-2016}.
We also exploit BPP heuristics to obtain three different greedy approaches for the HA-VNFP: \emph{Best Availability}, \emph{Best Fit}, and \emph{First Fit} greedy heuristics.
The algorithm, reported in \cref{algo:greedy}, starts from an empty initial placement and for each request $r$ it looks for a server having enough residual capacity to satisfy the demand $d_r$. If such a server is found, then the request is assigned to it, otherwise the algorithm fails without finding a feasible solution. 
However, we can observe that:
\begin{observation}
When $\sum_{r \in R} d_r \le \sum_{s `in S} q_s$ and split is allowed, our greedy heuristic always finds a feasible solution.
\end{observation}
In fact we can always split a request between two servers, as stated also in \cref{theo:feasibility-split}.

The selection of the server is performed by the procedure $\Call{selectServer}{\bar S, \bar d, split}$
which discards the servers without sufficient resources to satisfy demand $\bar d$, and 
selects a server depending on the chosen policy:
\begin{itemize}
\item \emph{best fit}: the server whose capacity best fits the demand;
\item \emph{first fit}: the first server found;
\item \emph{best availability}: the server with the highest availability.
\end{itemize}
While the first three policies are well-know for the BPP, the fourth one is designed for the HA-VNFP.

Master VNFs are placed during the assignment of the requests.
Then, in a similar way, the algorithm places additional slaves: for each master the algorithm looks for a server having enough capacity for a slave still using $\Call{selectServer}{}$ procedure.
After a server is found, the slave is placed. Such a procedure is repeated until no additional slave is placed.

\begin{algorithm}[ht]
\caption{Greedy heuristic procedure}
\label{algo:greedy}
{\small
\begin{algorithmic}[1]
\Function{greedy}{$R, S, split$}
\State $placement \gets \emptyset$
\ForAll{$r \in R \mid d_r >0$}
	\Comment Assignment of requests
	\If{$\exists \hat s \gets $\Call{selectServer}{$d_r, S, split$}}
		\State create VNF $f_r$ if it does not exists in $placement$
		\State assign request $r$ to server $\hat s$ in $placement$
		\State demand $d_r$ is decreased
	\Else
		\State \Return infeasible
	\EndIf
\EndFor
\Do
	\Comment Add slaves
	\ForAll{VNFs $v  \in placement$}
		\State $\bar S \gets $ servers of $S$ without $v$ and its slaves
		\If{$\exists \hat s \gets $\Call{selectServer}{$d_v, \bar S, FALSE$}}
			\State create slave of VNF $v$ on server $\hat s$ in $placement$
		\EndIf
	\EndFor
\While{slaves are found}
\State \Return $placement$
\EndFunction
\end{algorithmic}}
\end{algorithm}

\subsection{Variable Neighborhood Search}

The \emph{Variable Neighborhood Search (VNS)} is a meta-heuristic that systematically changes the neighborhood within the local search algorithm, in order to escape from local optima. In other words, it starts from an initial solution, applies a local search algorithm until it improves, and then changes the type of local search algorithm applied to change the neighborhood.
Our VNS algorithm explores $4$ different neighborhoods and it is initialized with several starting points, each obtained using a different greedy algorithm.

\begin{algorithm}[ht]
\caption{Variable Neighborhood Search}
\label{algo:vns}
{
\small
\begin{algorithmic}[1]
\Function{vns}{$R, S, split$}
\State $startingPoints \gets \{ $
\Call{bestAvailability}{$R, S, split$}, \linebreak  
\Call{bestFit}{$R, S, split$}, 
\Call{firstFit}{$R, S, split$} $\}$
\State $operators \gets \{vnfSwap, slaveSwap,requestSwap,\hfill$ \linebreak $requestMove\}$
\State $bestPlacement \gets \emptyset$
\ForAll{$placement \in startingPoints$}
\Do  
	\ForAll{$op \in operators$}
		\State $placement \gets$ apply $op$ to $placement$ 
		\If{$placement$ improves $ bestPlacement$}
			\State $bestPlacement \gets placement$
			\State \textbf{break}
		\EndIf
	\EndFor
\While{improving $bestPlacement$}
\EndFor
\State \Return $bestPlacement$
\EndFunction
\end{algorithmic}
}
\end{algorithm}

The main logic of our VNS algorithm is sketched in \cref{algo:vns}:
we generate $3$ starting points by using the greedy heuristics of \cref{sec:greedy} and 
we explore their neighborhood for a placement improving the availability. If no improvement can be found, the algorithm switches the neighborhood.

Indeed applying local search is time expensive, but we can observe that a max-min objective function divides the requests in two sets: a set of requests having an availability equal to the objective function and another set having a better availability. We refer to the former as the set of the \emph{worst requests}, since they are the ones whose improvement will also improve the availability of the entire solution. 
To reduce the computing time and focus our algorithm we found to be profitable to restrict the explored neighborhood to the worst requests only.
Also, after applying each operator we look for new slaves, as in the greedy procedure \cref{algo:greedy}.

Given two feasible placements, we say that one is improving if it has a higher availability or if it has the same availability but fewer worst requests.

In the following we describe the neighborhoods of our VNS.

\paragraph{VNFs swap}

\begin{figure}[ht]
\centering
\includegraphics[width=0.4\columnwidth]{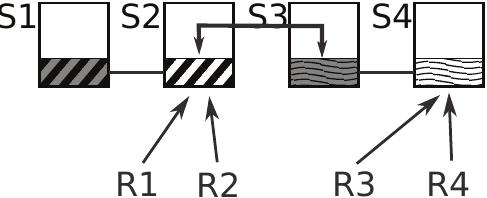}
\includegraphics[width=0.15\columnwidth]{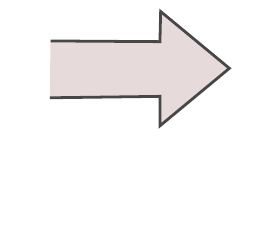}
\includegraphics[width=0.4\columnwidth]{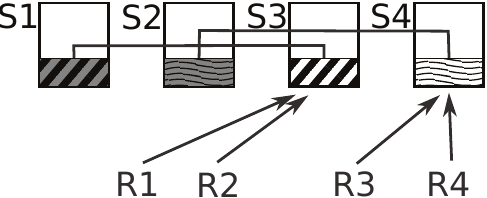}
\caption{Example of VNFs swap neighborhood: VNFs on server $S2$ and $S3$ are swapped. If a VNF is a master, then all its assigned requests are redirected to the new server.}
\label{fig:vnf-swap-n}
\end{figure}

The first neighborhood consists of swapping VNFs (see \cref{fig:vnf-swap-n}): given a VNF, we swap it with a subset of VNFs deployed on a different server. If the placement is improved, then we store the result as the best local change. 

In general our operator is $O(2^{|F| \cdot |S|})$ but we found profitable to set an upper bound of $1$ to the cardinality of the set of swapped VNFs, obtaining a $O(|F| \cdot |S|)$ operator.

\paragraph{Slave VNFs swap}

\begin{figure}[ht]%
\centering%
\includegraphics[width=0.4\columnwidth]{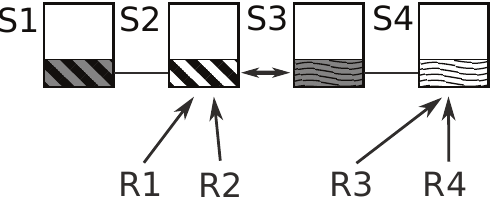}%
\includegraphics[width=0.15\columnwidth]{img/arrow}%
\includegraphics[width=0.4\columnwidth]{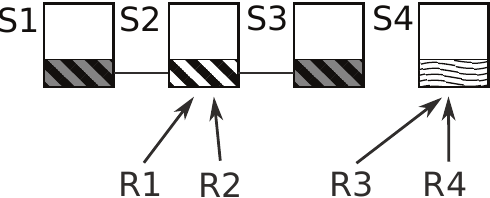}%
\caption{Example of slave VNFs swap neighborhood: a slave is removed from the placement in order to free resources for a slave of a different master VNF.}%
\label{fig:b-swap-n}
\end{figure}

We explore the neighborhood where a slave VNF is removed to free resources for an additional slave of a different master VNF (see \cref{fig:b-swap-n}). The complexity of this operator is $O(|F| \cdot |S|)$.

\begin{figure}[ht]
\centering
\includegraphics[width=0.4\columnwidth]{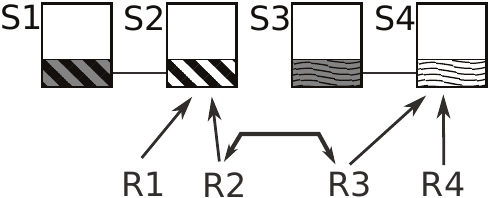}
\includegraphics[width=0.15\columnwidth]{img/arrow}
\includegraphics[width=0.4\columnwidth]{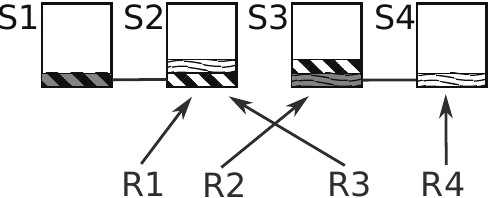}
\caption{Example of requests swap neighborhood: requests $R2$ and $R3$ are swapped changing the respective servers. When swapping a request, a new VNF instance is created if none existed on the new server. }
\label{fig:r-swap-n}
\end{figure}

\paragraph{Requests swap}
We also explore the neighborhood where requests are swapped (see \cref{fig:r-swap-n}): given a request we consider a subset of requests assigned to a different server and then swap the former with the latter.
Similarly to the swap of VNFs, the complexity of this operator is $O(2^{|R|})$. However, by setting an upper bound of $1$ to the cardinality of the swapped requests set we obtain a $O(|R|)$ operator.

\paragraph{Request move}

\begin{figure}[ht]
\centering
\includegraphics[width=0.4\columnwidth]{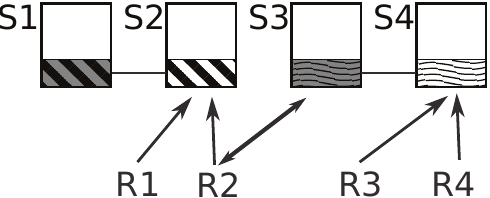}
\includegraphics[width=0.15\columnwidth]{img/arrow}
\includegraphics[width=0.4\columnwidth]{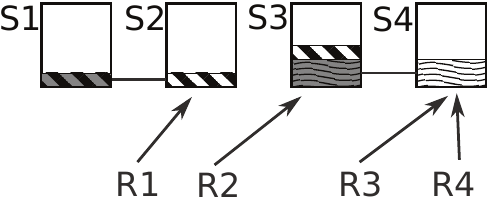}
\caption{Example of request move neighborhood: request $R2$ is moved and assigned to a different server. }
\label{fig:r-move-n}
\end{figure}

In the last exploration we consider the neighbors where a request is simply moved to a different server (see \cref{fig:r-move-n}). The complexity of this operator is $O(|S|)$.

In principle, even if all the operators polynomial time  our VNS algorithm is not . 
However, an upper bound $k$ to the number of iterations can be set, obtaining a $O(k\cdot |R| \cdot |S| \cdot  \max \{|R|, |F| \cdot |S|\})$ heuristic. Also, in the following we show that our VNS requires small computing time for NFVI of limited size and it can be parameterized to end within a time limit, making it suitable for both online and offline planning.

\section{Simulation}\label{sec:simulation}
We evaluate empirically the quality of our methodologies: the greedy heuristic using four different policies (best fit, first fit, and best availability), the VNS algorithm, and the mathematical programming formulation as a MIP. 
However we could run our MIP only on small instances with $3$ or $4$ servers and $50$ requests.
In our framework we first run the algorithms using the demand load balancing policy, and allowing split only if the former fails to assign all the requests.
All methodologies are implemented in C++, while CPLEX 12.6.3 \cite{cplex} is used to solve the MIP. The simulations have been conducted on Intel Xeon X5690 at 3.47 GHz. 
We also produced a graphical DSS tool integrating the VNS and the greed algorithms (in python) working on arbitrary 2-hop topologies and made it available in~\cite{HANFVsw}.

\begin{figure*}[ht]
\centering%
\subfigure[\label{fig:test-time-random}Average computing time (vertical axis in log. scale).]{\includegraphics[width=1.15\columnwidth]{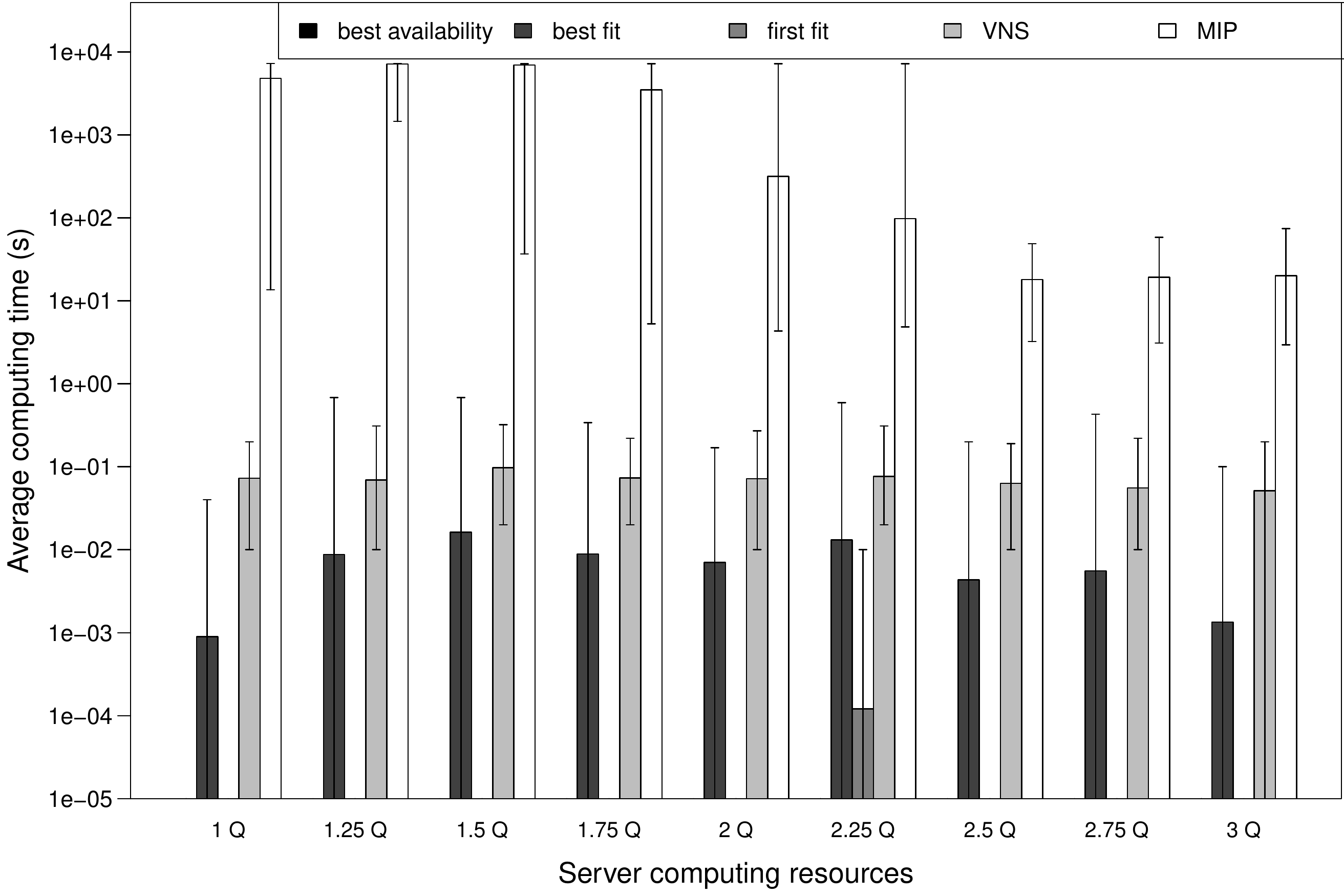}}%
\hfill%
\subfigure[\label{fig:test-availability-50}Average minimum availability.
]{\includegraphics[width=0.77\columnwidth]{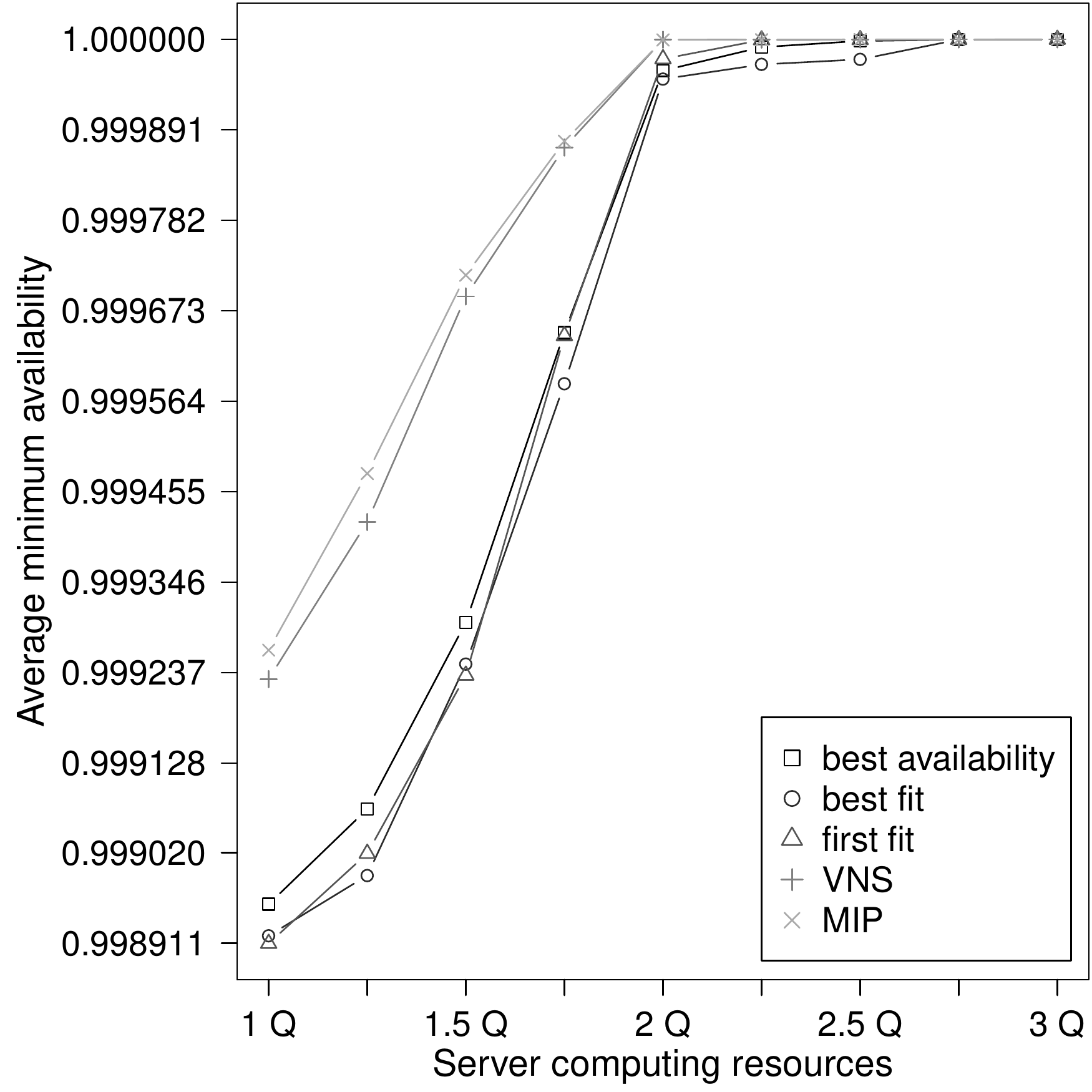}}%
\caption{Results for instances with $50$ requests.}
\end{figure*}

\subsubsection{Dataset generation}

We generated a random dataset consisting of instances that differ for the number of requests, total amount of computing resources, and availabilities of the network components.
We set the number of VNF types provided by our NFVI to $|F| =5$. We assumed an NFVI with $3$ clusters ($|C|=3$) and $3$ access points ($|P|=3$).
Each request has a random demand $d_r \in [1,10]$, while each server has a random capacity $q_s \in [75, 125]$.
The availabilities of all the components of our NFVI are selected between $\{0.9995, 0.9999, 0.99995, 0.99999\}$ as in \cite{Gill-failures, Yang2016, Zhang-2014}.

We generated $30$ instances for each combination of:
\begin{itemize}
\item number of requests $|R| = \{50, 100, 200, 300, 400, 500\}$;
\item number of access points for each request $|P_r| \in \{1, 2, 3\}$.
\end{itemize}

The number of servers depends on the number of requests, the total amount of the demands, and the random capacities: we generated a set of servers such that their capacities are enough to serve all the demands
$Q = \sum_{s \in S} q_s \ge \sum_{r \in R} d_r $. 
Note that such condition guarantees the feasibility only when splitting requests is allowed.
Servers are randomly distributed among all the clusters, in such a way that for each pair of clusters $c$ and $c'$ we have
$
|S_c - S_{c'}| \le 1. 
$
Under these conditions we obtained instances with around $3$ servers when $|R| = 50$ and $28$ servers when $|R| = 500$.

\subsection{Comparison on small instances}
We first evaluate the quality of our VNS heuristic against the solutions obtained by the MIP solver and the greedy heuristics.
In order to study how the NFVI behaves on different levels of congestion, we let its computing resources to grow from an initial value of $Q = \sum_{s \in S} q_s$, to $3 \cdot Q$, with a step of $0.25 \cdot Q$.
Due to the exponential number of the variables of our formulation, the MIP solver could handle only small instances with $50$ requests.
All tests have been performed setting a time limit of two hours each and all the algorithms managed to assign all requests without splits.

In \cref{fig:test-time-random} we show the average computing time of the algorithms: while the MIP hits several times the time limit, computing times are negligible for all the heuristics, and VNS can be considered as a good alternative for online orchestration when the set of the requests is small.
The optimization problem seems harder when the amount of computing resources is scarce: in fact, the average computing time of the MIP is closer to the time limit when the overall capacity is less than $2\cdot Q$. Instead, with higher quantities of resources the MIP always find the optimal solution within the time limit.

\begin{figure*}[ht]%
\centering%
\subfigure[\label{fig:availability-50-11}$1$ access point for request. ]{\includegraphics[width=0.3\textwidth]{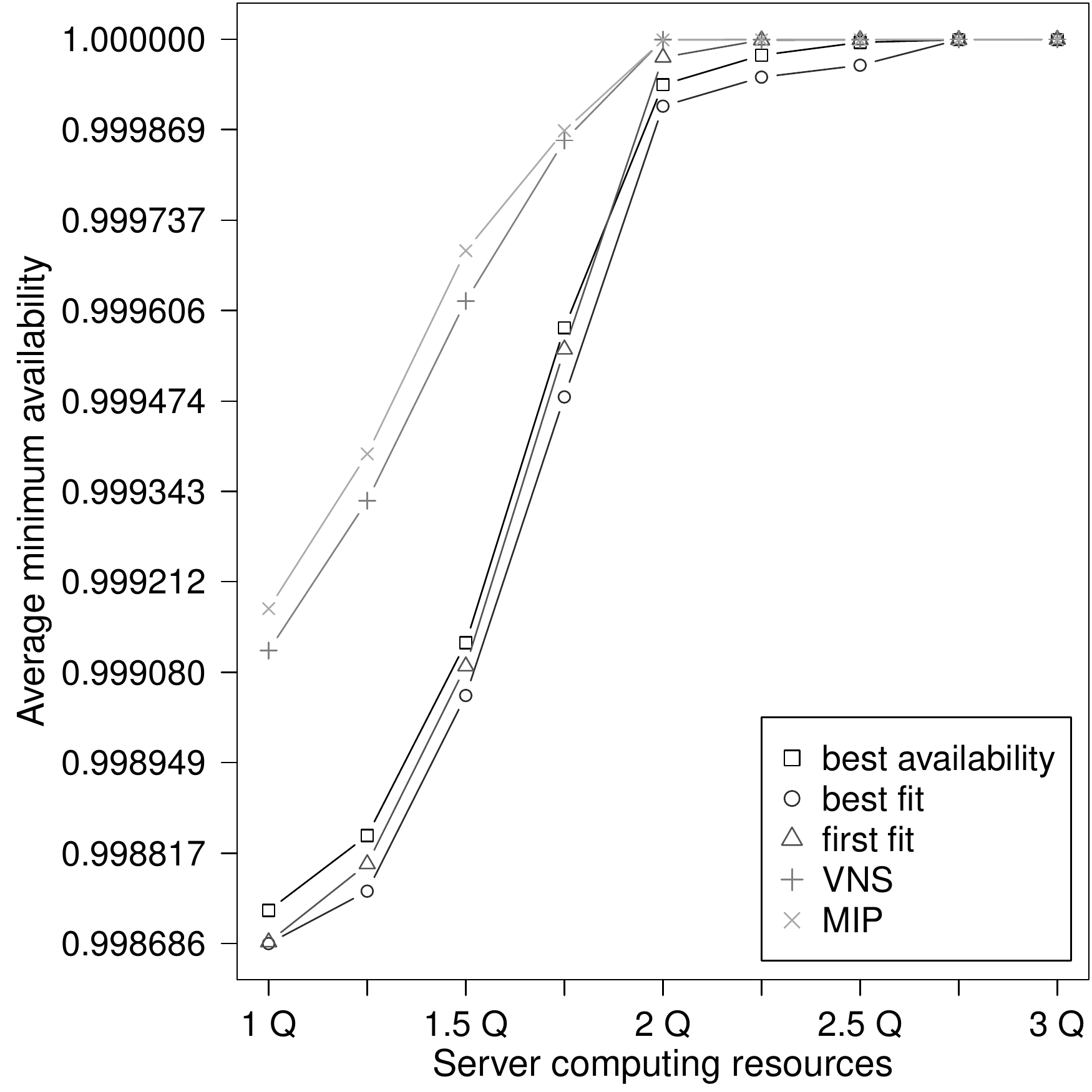}}%
\hfill%
\subfigure[\label{fig:availability-50-22}$2$ access points for request. ]{\includegraphics[width=0.3\textwidth]{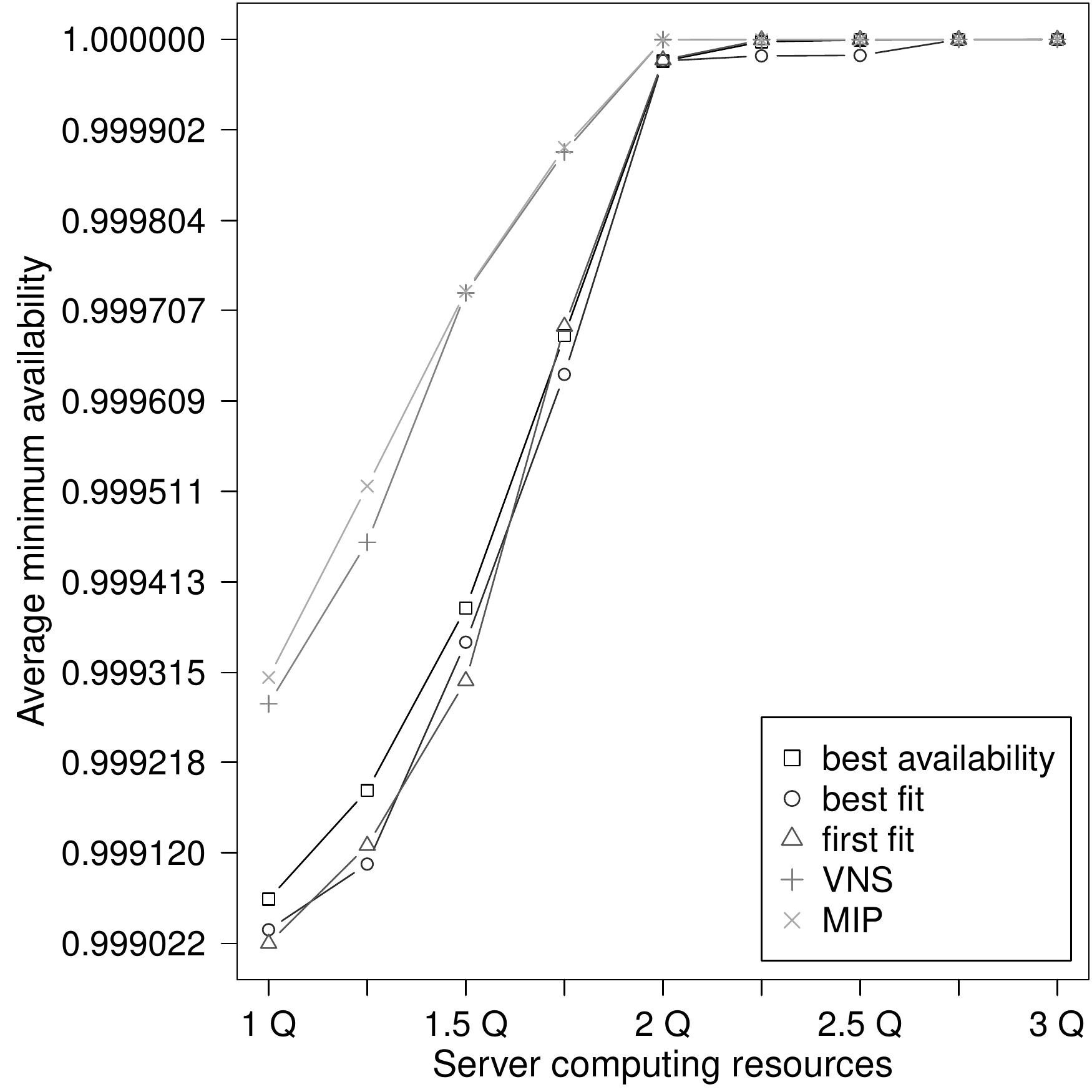}}%
\hfill%
\subfigure[\label{fig:availability-50-33}$3$ access points for request. ]{\includegraphics[width=0.3\textwidth]{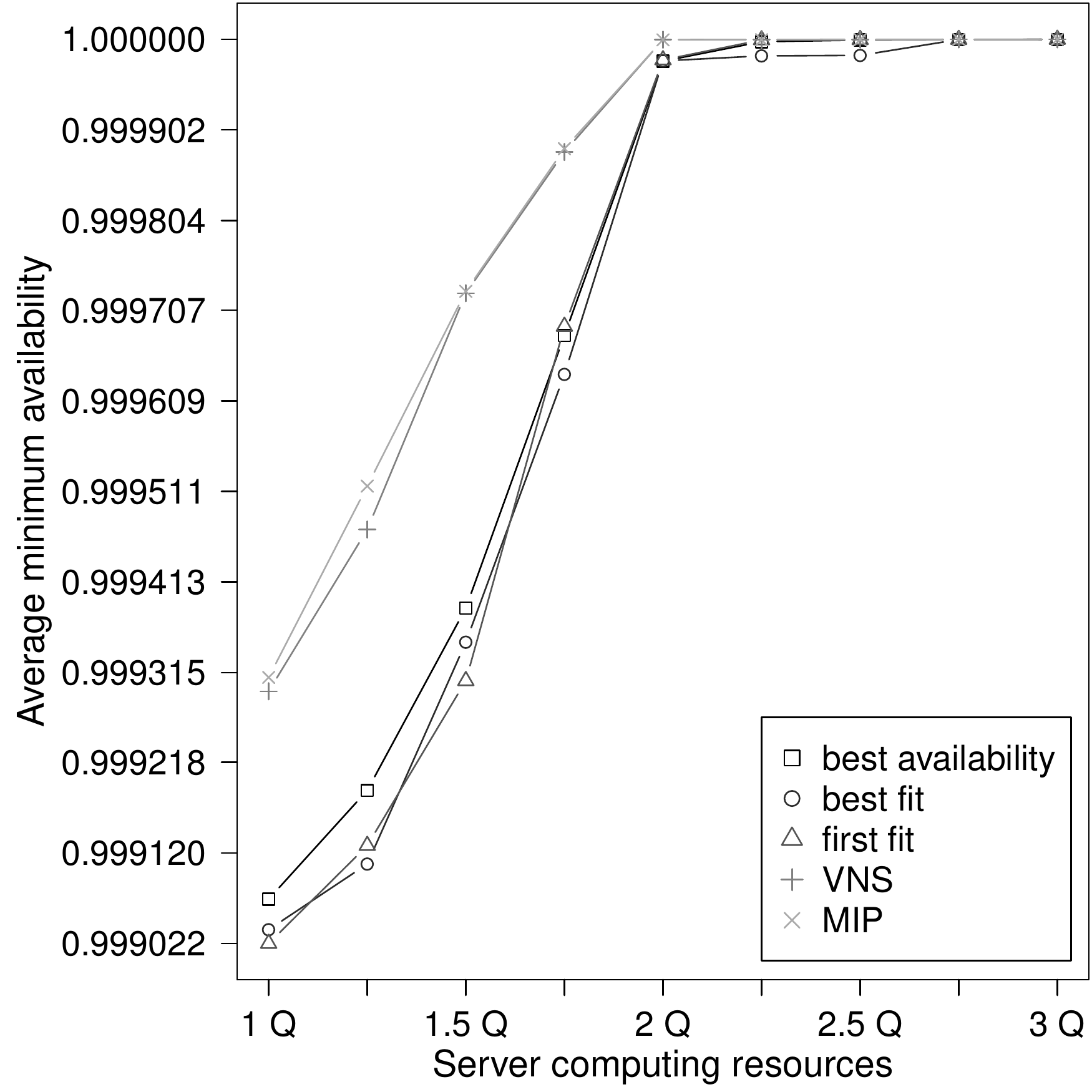}}%
\caption{Average minimum availabilities for instances with $50$ requests and different number of access points for each request.}%
\label{fig:availability-50-aps}%
\end{figure*}

In \cref{fig:test-availability-50} we show that the results of our VNS heuristic are close to the MIP ones, while there is a significant gap between the latters and the greedy heuristic ones.
In fact, both the MIP and the VNS succeed in finding solutions with an availability of three nines even with scarce resources.
Eventually all the algorithms reach an high level of availability when computing resources are doubled.

In \cref{fig:availability-50-aps} we show the variation of the availability when the number of access points for each request increases: in \cref{fig:availability-50-11}, \cref{fig:availability-50-22}, and \cref{fig:availability-50-33} we report the average availability when requests can be routed to the NFVI using $1$, $2$, and $3$ access points, respectively.
The path protection obtained by using more than one access point substantially increases the level of the availability. However, having more than $2$ access points does not provide additional benefits.

\begin{figure*}[htp]
\centering%
\subfigure[\label{fig:test-time-big}Average computing time (vertical axis in log. scale).]{\includegraphics[width=1.15\columnwidth]{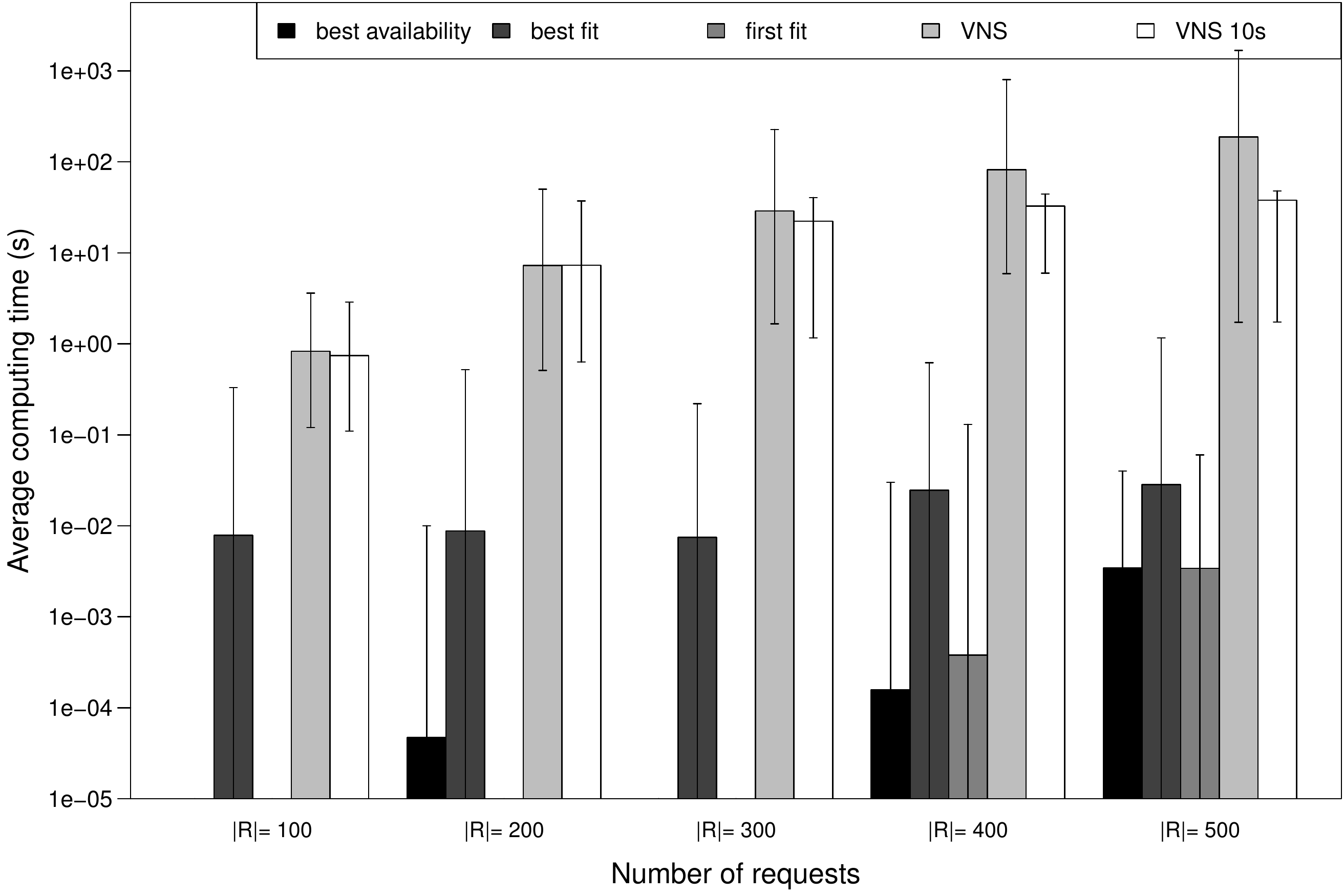}}%
\hfill%
\subfigure[\label{fig:test-availability-big}Average minimum availability.
]{\includegraphics[width=0.77\columnwidth]{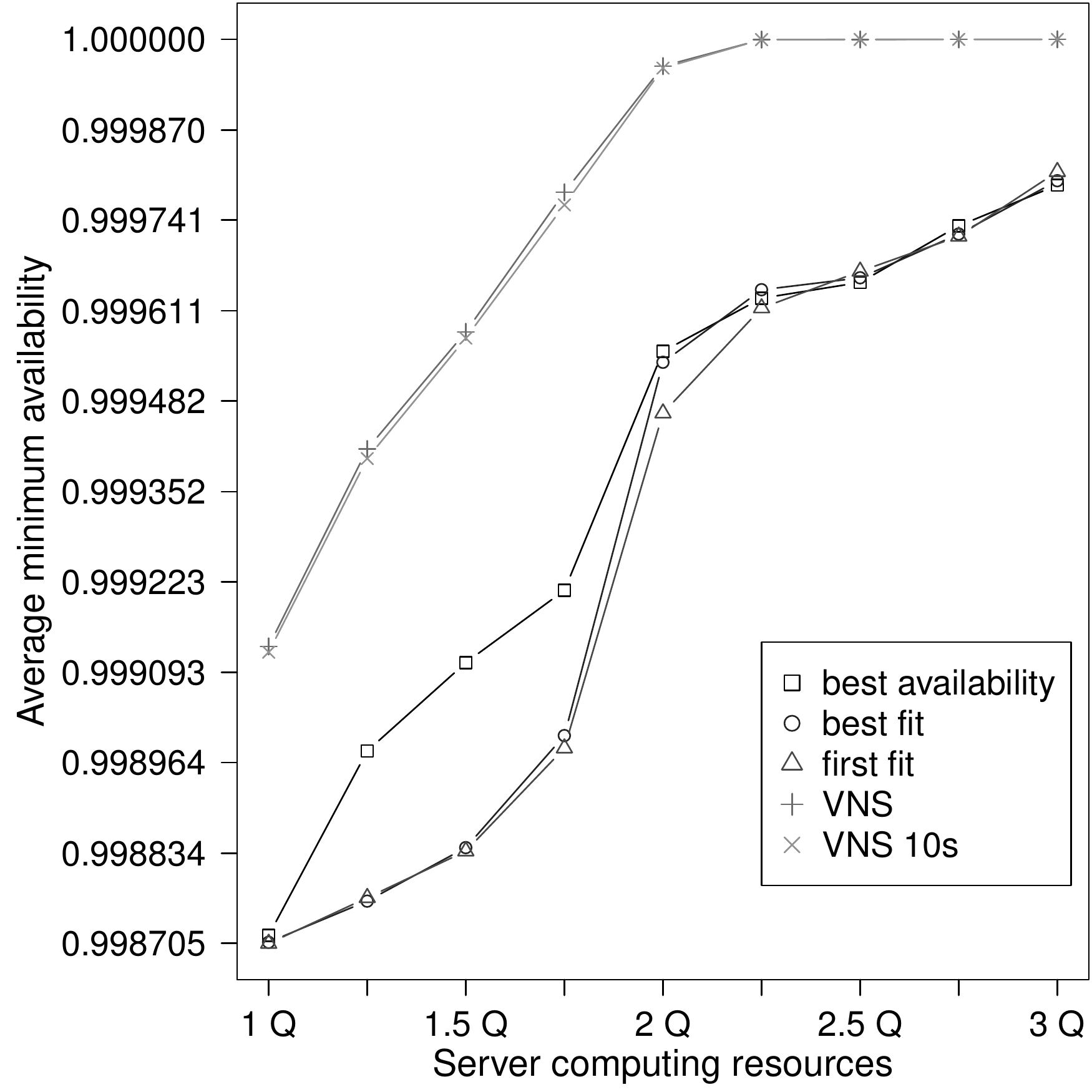}}%
\caption{Results for instances with up to $500$ requests.}
\end{figure*}

\begin{figure*}[htp]%
\centering%
\subfigure[\label{fig:availability-100}Instances with $100$ requests.]{\includegraphics[width=0.3\textwidth]{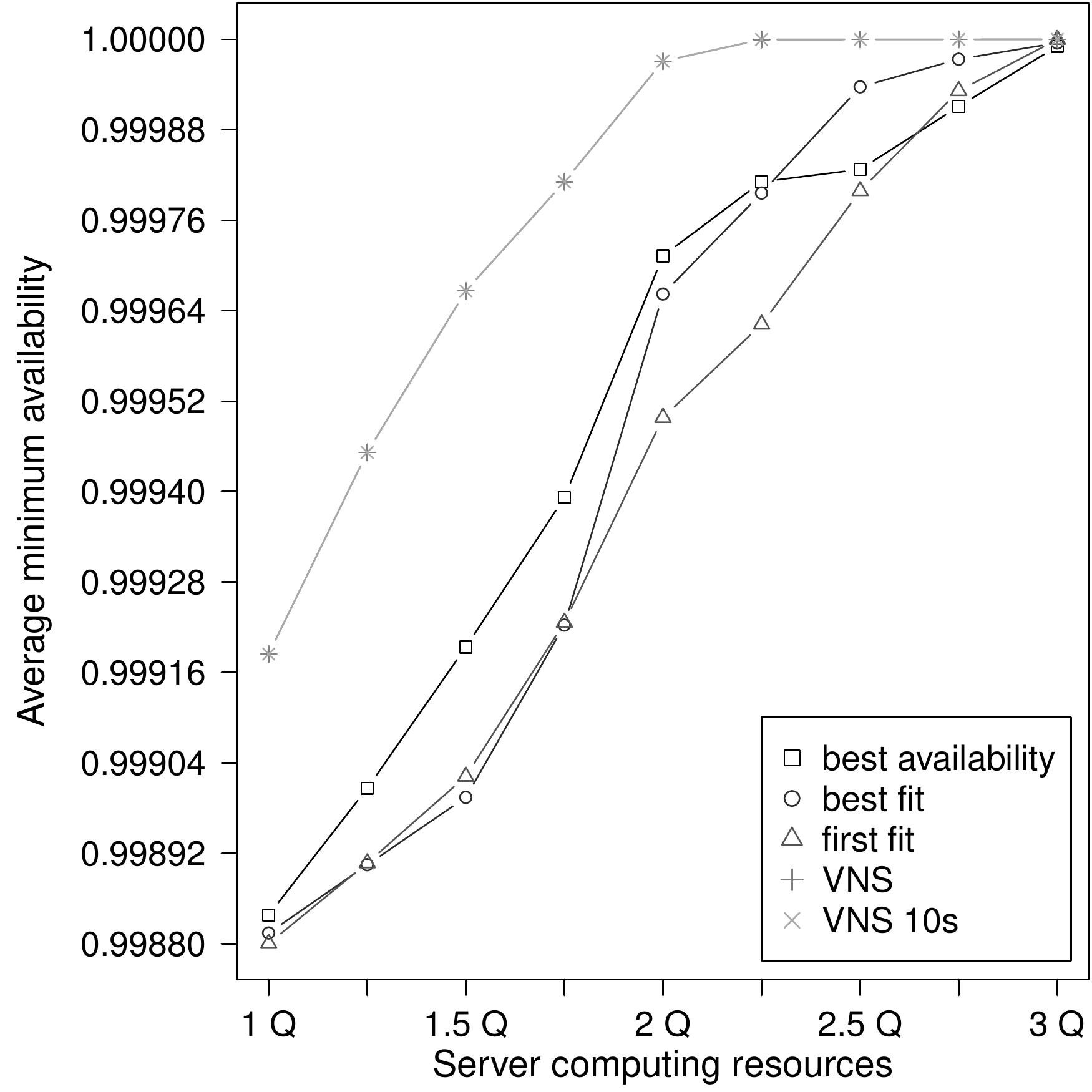}}%
\hfill%
\subfigure[\label{fig:availability-200}Instances with $200$ requests. ]{\includegraphics[width=0.3\textwidth]{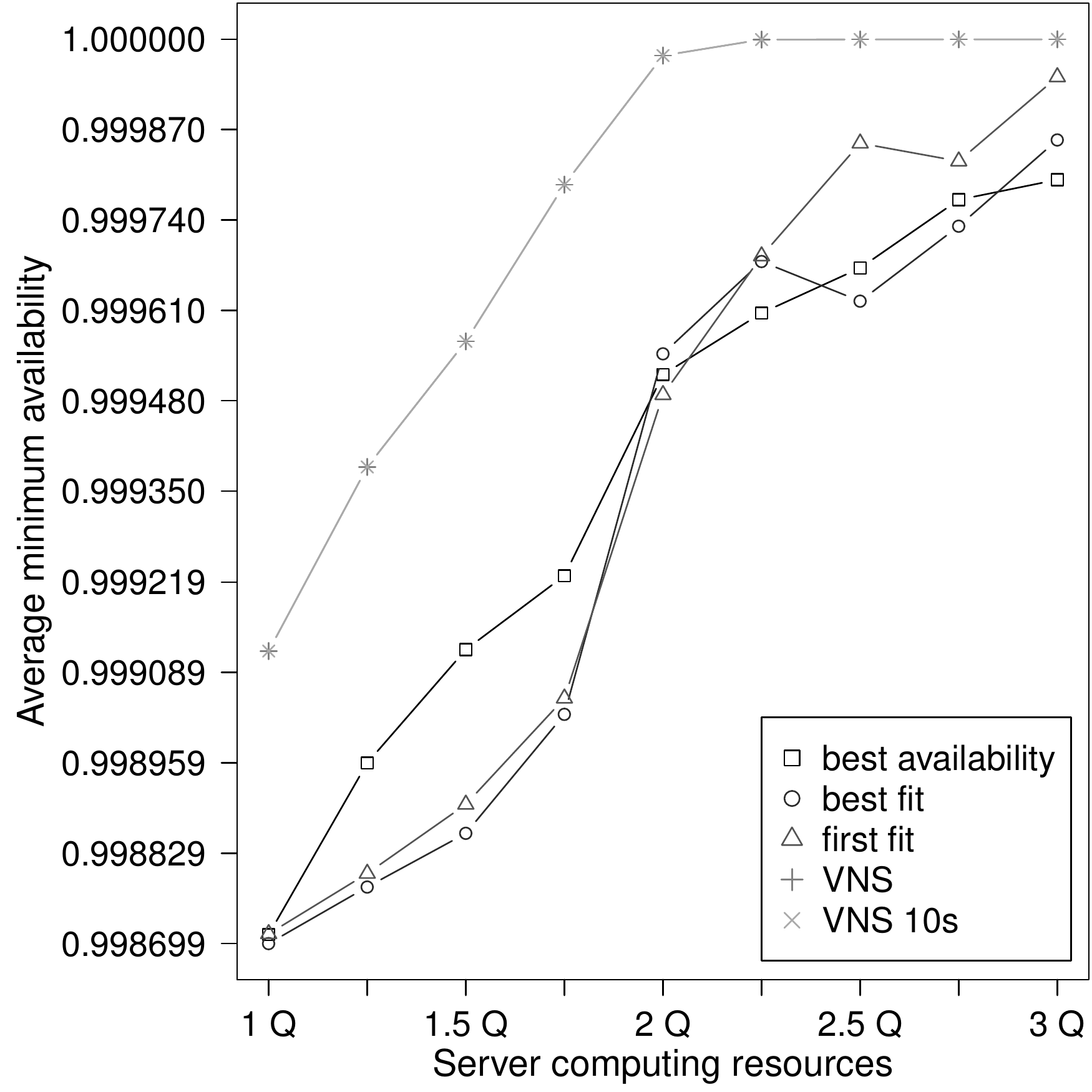}}%
\hfill%
\subfigure[\label{fig:availability-300}Instances with $300$ requests. ]{\includegraphics[width=0.3\textwidth]{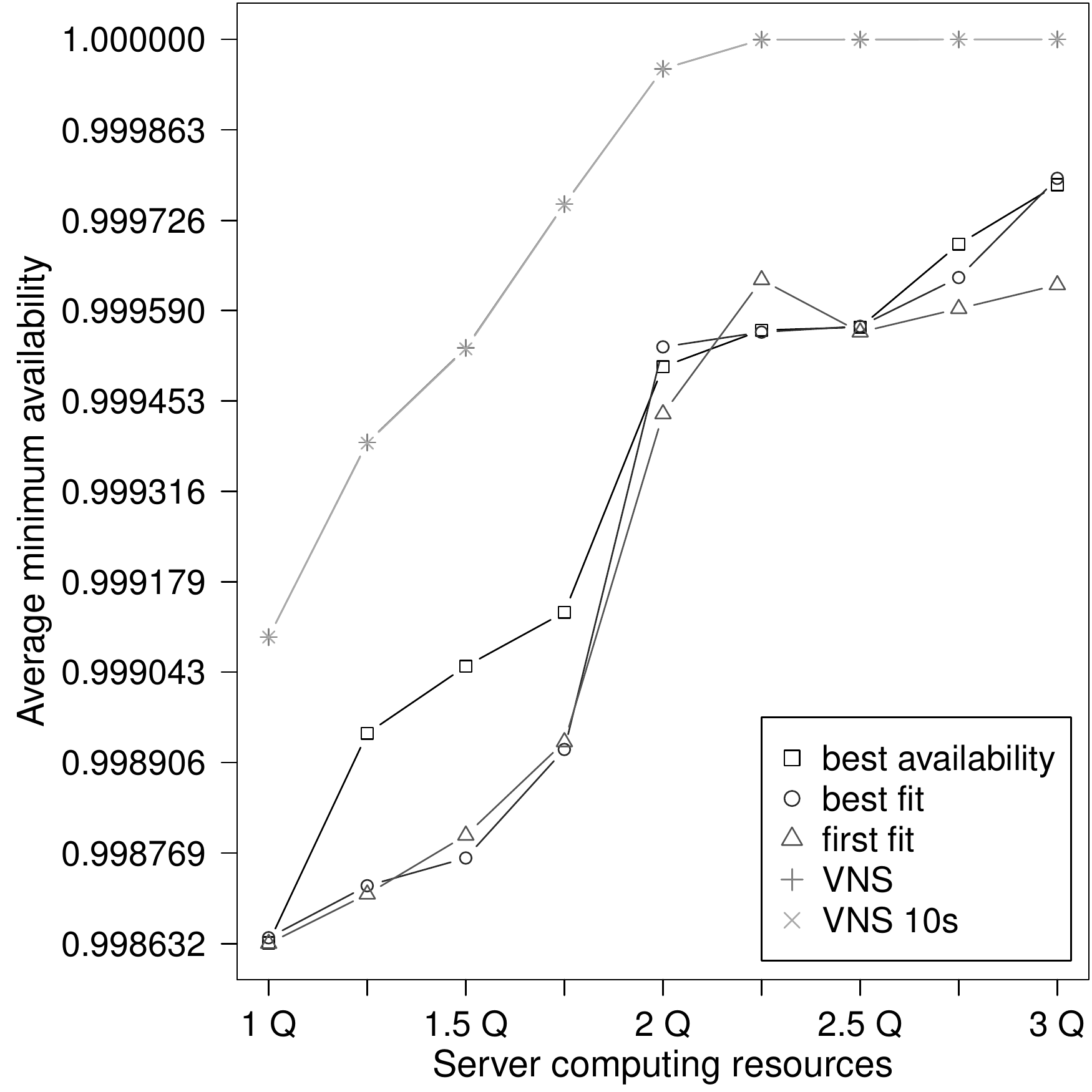}}%
\hfill
\subfigure[\label{fig:availability-400}Instances with $400$ requests. ]{\includegraphics[width=0.3\textwidth]{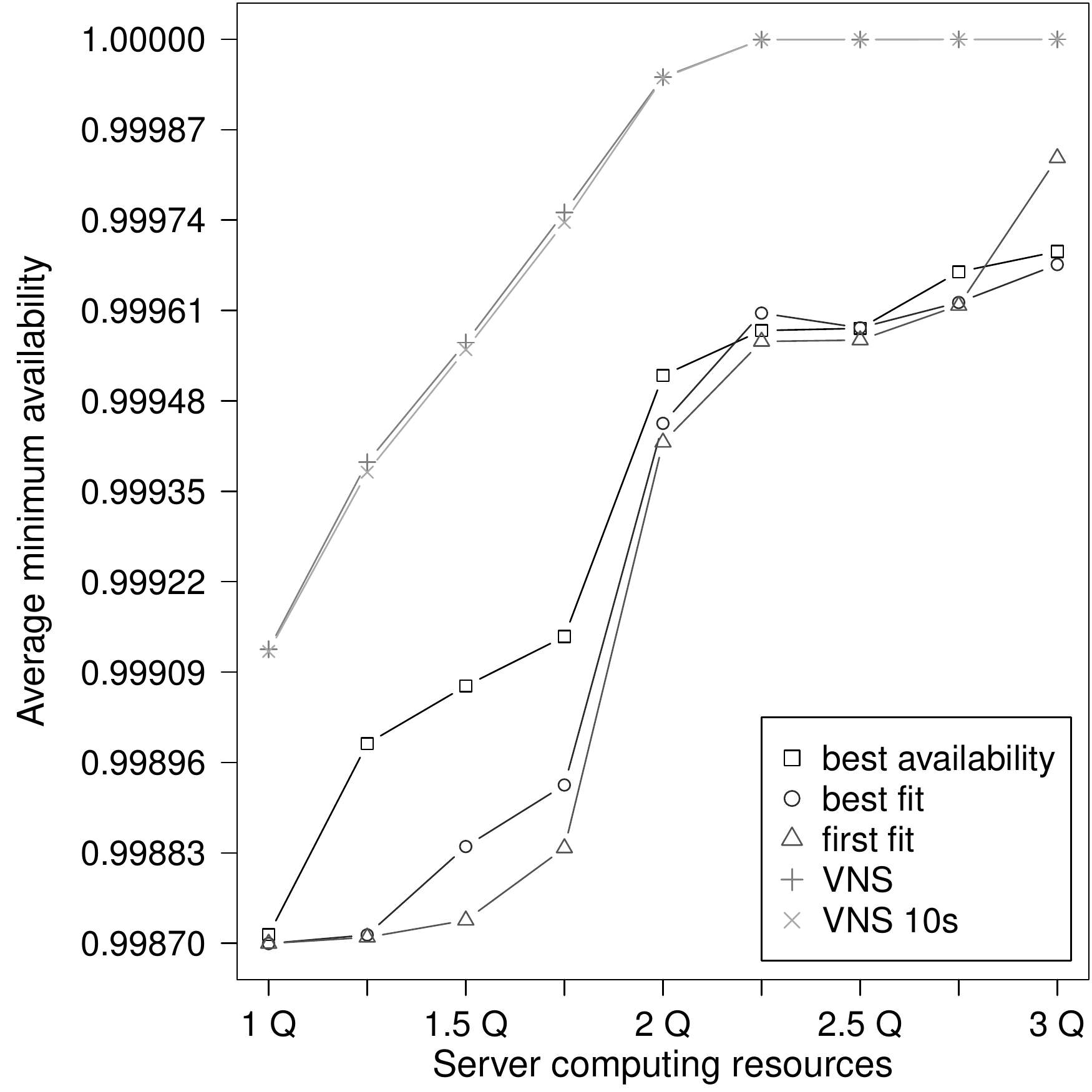}}%
\hfill%
\subfigure[\label{fig:availability-500}Instances with $500$ requests. ]{\includegraphics[width=0.3\textwidth]{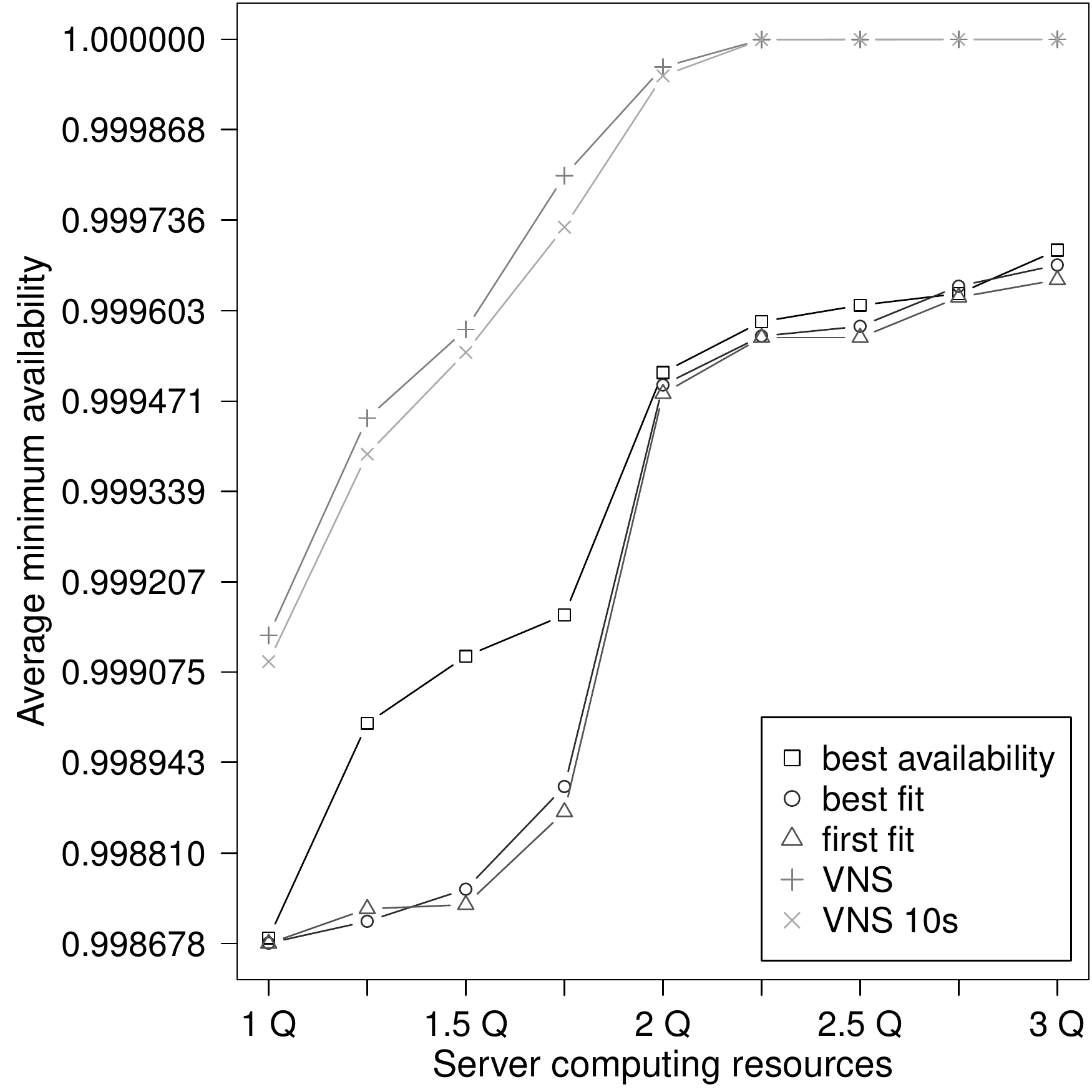}}%
\hfill%
\subfigure[\label{fig:availability-vns}VNS results only.
]{\includegraphics[width=0.3\textwidth]{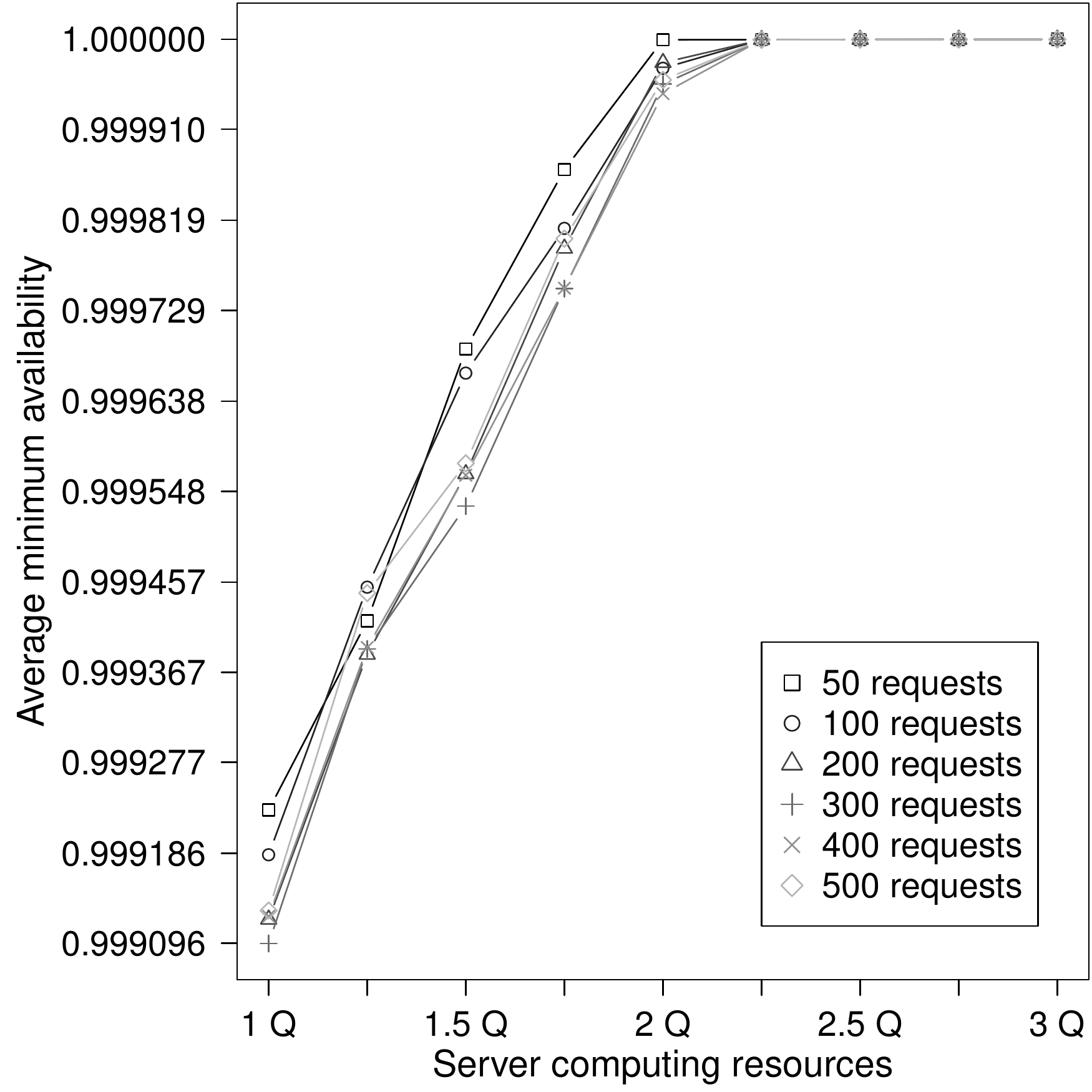}}%
\caption{Average minimum availabilities for instances with up to $500$ requests.}%
\vspace{-0.5cm}
\label{fig:availability-100-vns}%
\end{figure*}

\subsection{Scaling up the number of requests}
In a second round of experiments, we evaluated how our VNS algorithm behave when scaling up the number of requests.
However, when the number of servers increases its is not possible to use the MIP. Therefore, in the following analysis we compare the results of our VNS algorithm to the greedy ones only.
In addition, since our VNS algorithm has not polynomial time complexity, we include in the comparison the results obtained by setting a time limit of $10s$ at the exploration of each starting solution.

We can first observe in \cref{fig:test-time-big} that the computing time of the VNS algorithm grows exponentially when the number of requests increases. Indeed setting a time limit
reduces the overall computing time, which is always less than a minute. 

In \cref{fig:test-availability-big} we show that on average there is always a substantial gap between the results obtained by our VNS algorithms and the greedy heuristics. 
We can also observe that 
on average
the VNS time limit does not  penalize substantially the results.
Therefore our VNS algorithm with time limit can reduce computing times with minimal loss in availability.

From \cref{fig:availability-100} to \cref{fig:availability-500} we show the results on instances having from $100$ to $500$ requests individually. We can observe that the greedy heuristics progressively loose in quality of the solutions, and the gap with the VNS algorithms increases with the number of the requests. 

Finally in \cref{fig:availability-vns} we show only the results concerning the VNS algorithm without time limit and how it behaves when both the number of requests and the overall capacity increase.
We can observe that the curves are similar and the quality of the solutions provided by our algorithm is not affected by the increasing of the size of the instances. Our VNS algorithm always provides placements with an availability of three nines even when resources are scarce, and it always reaches four nines when the capacity is doubled.

\section{Conclusion}\label{sec:conclusion}

We defined and modeled the HA-VNFP, that is the problem of placing VNFs in NFVI guaranteeing high availability. We provided a quantitative model based on probabilistic approaches to offer an estimation of the availability of a NFVI.
We proved that the arising nonlinear optimization problem is $\mathcal{NP}$-hard and we modelled it by means of a linear formulation with an exponential number of variables. However, to solve instances with a realistic size we designed both efficient and effective heuristics. 
By extensive simulations, we show that our VNS algorithm
finds solution close to the MIP ones, but in computing times smaller of orders of magnitude. 
We highlighted the substantial gap between the availability obtained using classic greedy policies, and the one obtained with a more advanced VNS algorithm, when the NFVI is congested.
Our VNS algorithm showed to be a good compromise to solve HA-VNFP in reasonable computing time, proving to be a good alternative for both online and offline planning.

We integrated our HA-VNFP algorithms in a graphical simulator made available with tutorial videos in~\cite{HANFVsw}.

\section*{Acknowledgments}
{
This article is based upon work from COST Action CA15127 ("Resilient
communication services protecting end-user applications from disaster-based failures - RECODIS")
supported by COST (European Cooperation in Science and Technology). This work was funded by the ANR Reflexion (contract nb: ANR-14-CE28-0019) and the FED4PMR projects.
}

\end{document}